\documentclass[a4paper, amsfonts, amssymb, amsmath, reprint, showkeys, nofootinbib, twoside, hidelinks, superscriptaddress]{revtex4-2}
\usepackage[english]{babel}
 \addto\captionsenglish{} 
\usepackage[utf8]{inputenc}
\usepackage[colorinlistoftodos, color=green!40, prependcaption]{todonotes}
\usepackage{amsthm}
\usepackage{mathtools}
\usepackage{physics}
\usepackage{xcolor}
\usepackage{graphicx}
\usepackage[left=23mm,right=13mm,top=35mm,columnsep=15pt]{geometry} 
\usepackage{adjustbox}
\usepackage{placeins}
\usepackage[T1]{fontenc}
\usepackage{lipsum}
\usepackage{csquotes}
\usepackage{geometry}

\usepackage{graphicx}
 \usepackage{wrapfig} 
 \usepackage{float} 
 \usepackage{adjustbox}
\usepackage{amsmath}
 \usepackage{bm}
\usepackage[version=3]{mhchem} 

 \usepackage{ amssymb }
\usepackage{MnSymbol}
 \usepackage{esint}
 \usepackage{gensymb} 
 \usepackage{braket}
 \usepackage{empheq}

 \usepackage{setspace}

 \usepackage[colorlinks=true]{hyperref}
\hypersetup{%
    ,urlcolor=blue
    ,citecolor=blue
    ,linkcolor=blue
    ,citecolor=blue
    ,allcolors=blue
    }

\begin{document}
\title{Magnetotransport in a graphite cylinder under quantizing fields}

\author{N. Kunchur}
\address{Max Planck Institute for Chemical Physics of Solids, Nöthnitzer Straße 40, 01187 Dresden, Germany}
\author{S. Galeski}
\address{Max Planck Institute for Chemical Physics of Solids, Nöthnitzer Straße 40, 01187 Dresden, Germany}
\address{Physikalisches Institut, Universität Bonn, Nussallee 12, 53115 Bonn, Germany}
\author{F. Menges}
\address{Max Planck Institute for Chemical Physics of Solids, Nöthnitzer Straße 40, 01187 Dresden, Germany}
\author{R. Wawrzyńczak}
\address{Max Planck Institute for Chemical Physics of Solids, Nöthnitzer Straße 40, 01187 Dresden, Germany}
\author{C. Felser}
\address{Max Planck Institute for Chemical Physics of Solids, Nöthnitzer Straße 40, 01187 Dresden, Germany}
\author{T. Meng}
\address{Institute of Theoretical Physics and W\"urzburg-Dresden Cluster of Excellence ct.qmat, Technische Universit\"at Dresden, 01062 Dresden, Germany}
\author{J. Gooth}
\address{Max Planck Institute for Chemical Physics of Solids, Nöthnitzer Straße 40, 01187 Dresden, Germany}
\address{Physikalisches Institut, Universität Bonn, Nussallee 12, 53115 Bonn, Germany}


\begin{abstract}
We analyze the transport properties of curved, three-dimensional graphite samples in strong magnetic fields. Focusing on a millimeter-scale graphite
cylinder as a prototypical curved object, we perform longitudinal and Hall voltage measurements while applying quantizing magnetic fields. These measurements are investigated as a function of field strength and angles. Most importantly, we find that angle-dependent Shubnikov-de Hass oscillations are superimposed with angle-independent features. Reproducing the experimental observations, we introduce a network model that accounts for the cylindrical geometry effect by conceptualizing the cylinder as composed of strips of planar graphite in an effectively inhomogeneous magnetic field. Our work highlights how the interplay between geometric curvature and quantizing magnetic fields can be leveraged to engineer tunable spatial current densities within solid-state systems, and paves the way for understanding transport properties of curved and bent three-dimensional samples more generally.
\end{abstract}

\maketitle

\section{Introduction}
In recent years, the analysis of transport properties of curved, bent, and twisted materials has become increasingly important~\cite{gentile2022electronic,nigge2019room,rogers2011synthesis,carr2020electronic,torma2022superconductivity}. Whereas considerable progress has been made for two-dimensional and layered materials~\cite{lorke2003curved,ferrari2008cylindrical,leadbeater1995electron,mendach2006evenly,vorob2007giant,leadbeater1995magnetotransport,vorobyova2015magnetotransport,friedland2007measurements,grayson2004quantum}, much less is known about the transport properties of bulk three-dimensional curved objects. This is in particular true in the presence of magnetic fields.  In this work, we provide a basic framework for understanding magnetotransport in three-dimensional curved samples. As a controlled reference geometry, we focus on three-dimensional cylinders, and provide an in-depth analysis of their experimentally observed transport behavior. Based thereon, we develop a simple phenomenological model that reproduces the observed data. Because the model can straight-forwardly be adapted to other geometries, our findings pave the way for magnetotransport studies of curved three-dimensional systems more generally.

As we will discuss below, a central ingredient in both the experimentally observed magnetotransport and the theoretical model is a non-uniform and tunable distribution of current across the sample. This non-trivial current distribution results from the interplay of sample geometry and applied magnetic field. The surface normal of a curved sample provides a local reference vector with which the direction of the applied field can be compared. Naturally, the relative orientation of these two vectors varies in space when curved samples are considered. As we will show below, the spatial dependence of the relative orientation of surface normal and magnetic field directly connects to a local variation of transport properties, which furthermore depends on the strength of the applied field. Quite generally, this means that the application of magnetic fields provides a convenient tool to imprint spatial sample geometry onto transport properties. 

This is to be contrasted with the behavior of a typical flat system. There, charge transport usually occurs via charge current paths dictated by the applied electric field and Ohm's law. The resistance is related to the resistivity merely via the length and the cross-section of the sample, and the electric field is parallel to the current path.  Applied magnetic fields (\textit{B}) introduce one layer of additional complexity: due to the Lorentz force, the conductivity becomes a tensor, and currents are in general not parallel to the electric fields in the sample.  

When the magnetic field becomes ``quantizing'', i.e. sufficiently strong that the splitting of energy levels into Landau levels becomes important for observable quantities, another layer of complexity is added. In the prominent case of the quantum Hall effect (QHE) in a two-dimensional electron gas (2DEG)~\cite{klitzing1980new}, the Hall resistivity undergoes quantization in units of $h/e^2$ while the longitudinal resistivity vanishes simultaneously. These features are typically explained by the existence of both localized and extended current-carrying states combined with topological band gaps~\cite{laughlin1981quantized}. A microscopic description in real space hints at the presence of charge carriers arranged along compressible and incompressible regions~\cite{weis2011metrology}. After investigating the QHE in two-dimensional (2D) materials like graphene~\cite{zhang2005experimental,novoselov2004electric}, the exploration of topological phases~\cite{hasan2010colloquium} has extended into three-dimensional (3D) realm~\cite{zhang2009topological,chen2009experimental,xu2015experimental,lv2015experimental}. Signatures of the QHE attributed to the magnetotransport in graphite~\cite{kempa2006integer} and the unexplored edge state physics in this material pose open questions. Additionally, the extension of QHE to a quasi-quantized Hall effect~\cite{gooth2023quantum, galeski2021origin,galeski2020unconventional, wawrzynczak2022quasi,manna2022three} in systems with closed 3D Fermi surfaces prompts further investigation, particularly into hypothesized surface current-carrying states~\cite{gooth2023quantum}. 

The above examples clearly illustrate the importance of understanding transport in quantizing fields in 3D system. As of now, however, non-trivial sample geometries have not played an important role. In this respect, the QHE in 2D samples can serve as a useful reference point: the impact of non-trivial sample geometries and curvature has been studied to some extent for Hall bars~\cite{lorke2003curved,chaplik1998effect}. A rather useful simple picture is that the local normal component of the applied homogeneous field dominates Hall physics. As a result, the Chern number~\cite{thouless1982quantized} (more precisely, the local Chern marker) can vary in space, which in turn results in topological transitions~\cite{thouless1982quantized} as a function of position. These transitions involve edge states carrying current that arise not at sample boundaries, but at the interfaces of gapped local ``Chern patches''. This simple picture explains the observed asymmetric magnetoresistance, a distinct feature of the magnetotransport in curved Hall bars caused by the specific current distribution~\cite{leadbeater1995magnetotransport,vorob2007giant,mendach2006evenly,friedland2009quantum,hugger2007influence}. Furthermore, by incorporating screening effects, the appearance of a metal-like compressible bulk at plateau transitions is highlighted, accounting for additional peaks in longitudinal transport~\cite{friedland2009quantum}. Moreover, beyond curved Hall systems, curved and bent 2D electronic systems have constituted a major theme of interest in recent years. In a constant magnetic field directed normally to its surface, localized cyclotron orbits are predicted to accompany snaking states at angular positions marking a change in the field gradient~\cite{chang2014strongly,ferrari2008cylindrical,bellucci2010landau,muller1992effect,manolescu2013snaking,chang2017theoretical,hugger2007influence}, resulting in transport signatures such as anisotropic magnetoresistance~\cite{chang2014strongly,chang2017theoretical}. 

While a substantial body of literature has thus analyzed the properties of curved 2D quantum Hall systems much less is known about the magentotransport properties of curved 3D samples. Those differ not only in dimensionality, but also in the fact that they typically do not exhibit a bulk gap when subject to magnetic fields due to a non-trivial dispersion along the direction parallel to the field. To analyze magnetotransport of curved 3D samples, we proceed as follows. In Sec.~\ref{section2}, the experimental procedure to investigate angle-resolved Hall and longitudinal magnetotransport is discussed. In Sec.~\ref{section3}, the key observation, namely the appearance of angle-independent features (AIF) in superposition with Shubnikov-de Hass (SdH) oscillations is summarized. Based on this observation, a network model is systematically developed, and the results from its simulation are discussed in Sec.~\ref{section4}, followed by the conclusions in Sec.~\ref{section5}.

\section{Experimental setup and planar reference transport}
\label{section2}

\begin{figure}[ht]
\includegraphics[width=\linewidth]{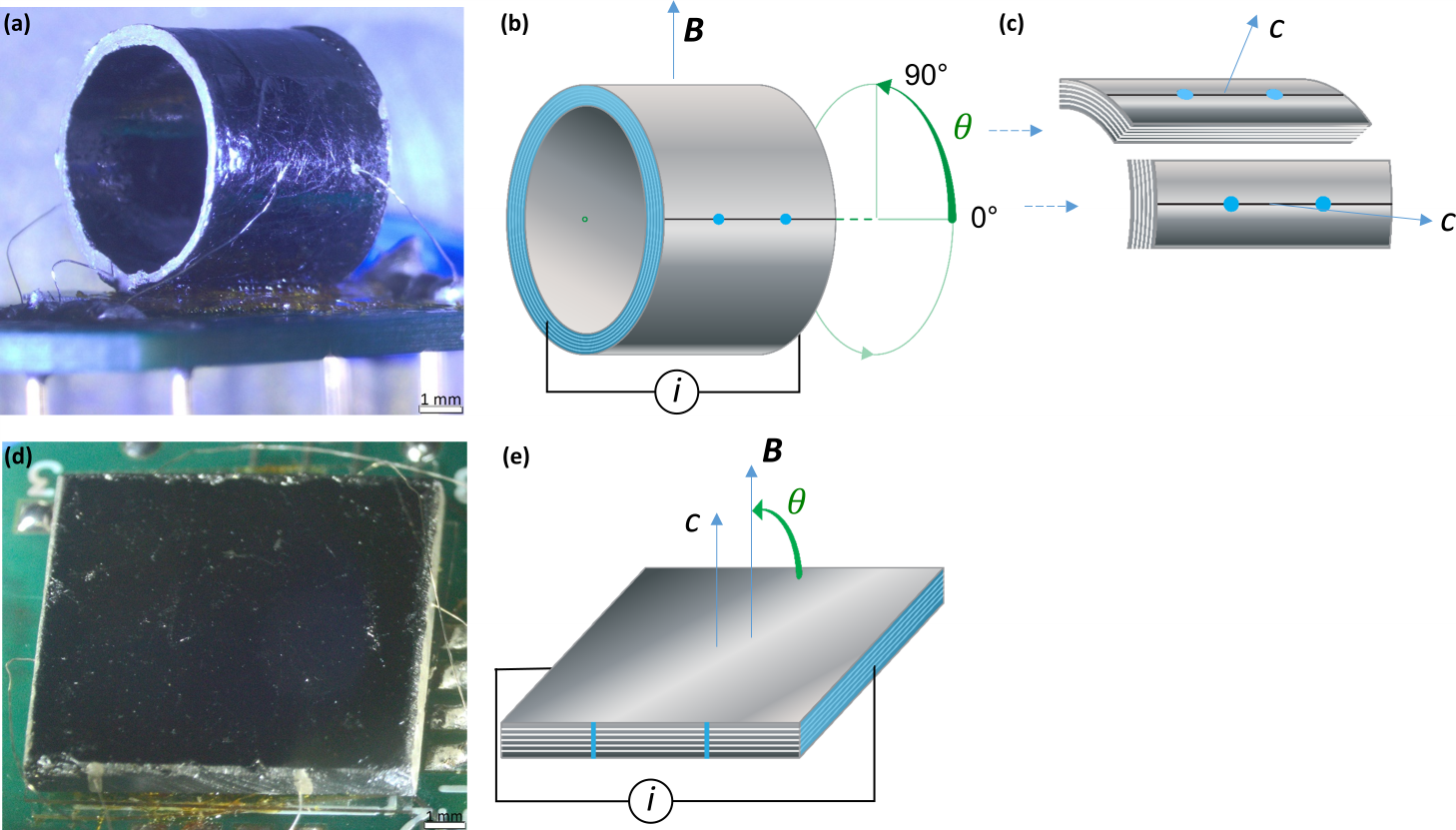}
\caption{(a) CHOPG sample with dimensions: radius = 3~mm, length = 5~mm and thickness = 0.3~mm. (b) Schematic of CHOPG with its axis marked by the dashed line. Longitudinal and Hall voltage are measured between contacts along the axis and across diametrically opposite ends, respectively. (c) Infinitesimal strips in CHOPG at $\theta$ (top) and 0$\degree$ (bottom), with their \textit{c} axes directed radially outward. (d) PHOPG sample with dimensions: 10~mm $\times$ 10~mm $\times$ 1~mm. (e) Current (\textit{i}) is sourced and drained across opposite faces. The longitudinal and Hall voltages are measured along and perpendicular to \textit{i}. PHOPG is rotated around the axis along \textit{i}. $\theta$ is measured between the plane normal to the \textit{c} axis of graphite and \textit{B}.}
\label{fig:Manuscript_FIG._1}
\end{figure}

Our experimental setup features a cylindrical sample of highly oriented pyrolytic graphite (HOPG), denoted as CHOPG (FIG.~\ref{fig:Manuscript_FIG._1}a). We select HOPG as our material of interest because it can be molded into various geometries by pressing polymers at high temperatures and pressures~\cite{freund2022optimization}. Despite not being single crystalline, HOPG shows clear signatures of Shubnikov-de Hass (SdH) oscillations, entering the quantum limit at moderate fields due to the small size of its 3D Fermi pockets~\cite{slonczewski1958band,mcclure1957band}. The current \textit{i} in CHOPG is sourced and drained uniformly across opposite circular rims. The interplay of \textit{B} directed along \textit{z} and the sample curvature can be investigated by rotating the cylinder around its axis. This shifts the pairs of contacts to the angular positions $\theta$ and $\theta+180\degree$, measured in the anti-clockwise direction (FIG.~\ref{fig:Manuscript_FIG._1}b and FIG.~\ref{fig:Manuscript_FIG._1}c). As a crucial reference for the interpretation of the obtained CHOPG data, similar longitudinal and Hall voltage measurements were also performed on planar samples of HOPG, denoted as PHOPG and shown in  FIG.~\ref{fig:Manuscript_FIG._1}d, using the standard configuration described in FIG.~\ref{fig:Manuscript_FIG._1}e.

\section{Magnetotransport measurements}
\label{section3}

\begin{figure}[ht]
\centering
\includegraphics[width=\linewidth]{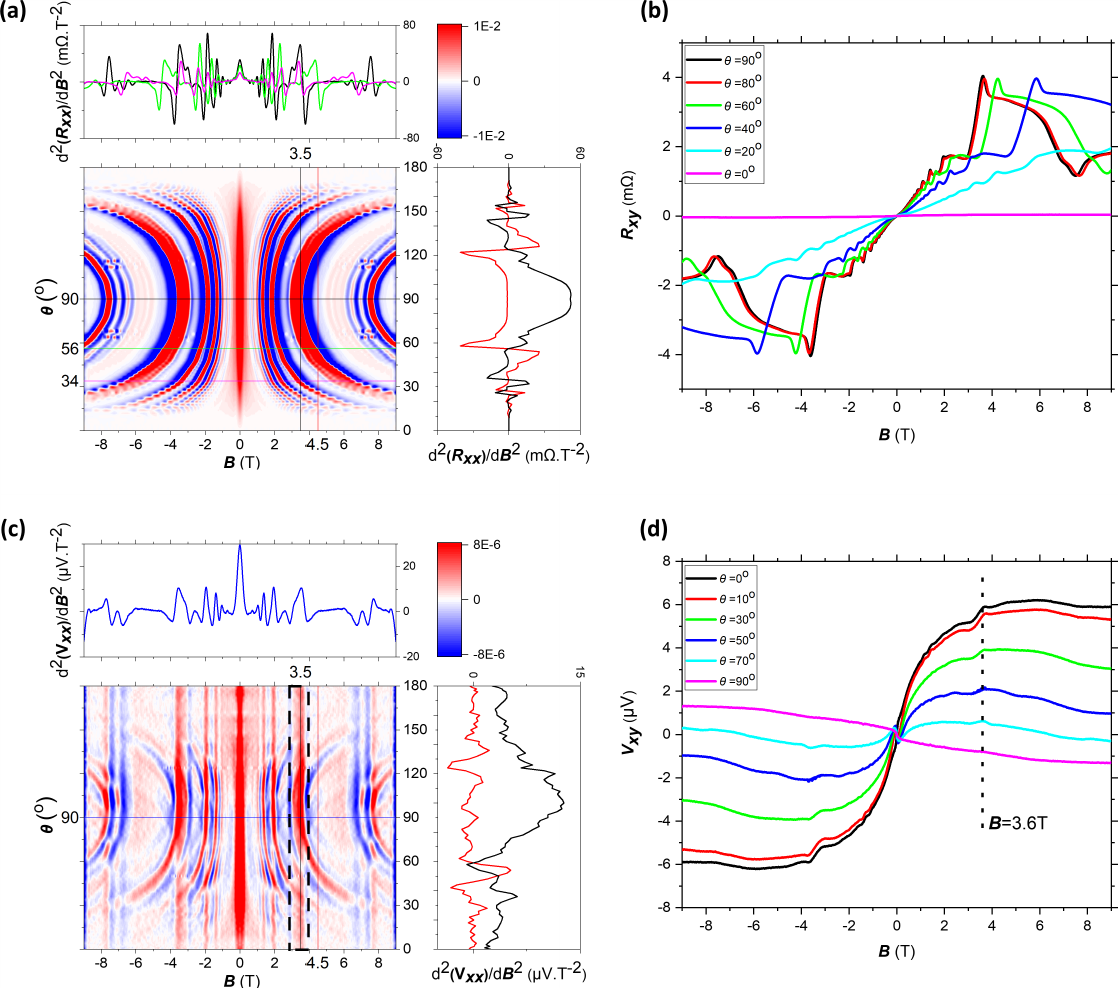}
\caption{(a) Contour plot of SdH oscillations in $d^2R_{xx,\theta}(B)/dB^2$. (b) $R_{xy}(B)$ in PHOPG at $\theta$. (c) Contour plot of oscillations in $d^2V_{xx,\theta}(B)/dB^2$. The SdH oscillations superimpose the AIF. An AIF at 3.5T is highlighted in the dashed black box. (d) $V_{xy}(B)$ in CHOPG showing pronounced features at identical values of $B$ for all $\theta$, highlighted by the dashed black line at $B=3.6$~T.}
\label{fig:Manuscript_FIG._2}
\end{figure}

To unravel the impact of sample curvature on magnetotransport, we performed a series of contrasting measurements in both PHOPG and CHOPG. For improved visibility, the SdH oscillations in PHOPG arising due to the Landau quantization are evaluated from the second derivative of $R_{xx,\theta}(B)$ with respect to \textit{B}. Here, $R_{xx,\theta}(B)$ represents the symmetric part of the magnetoresistance with respect to \textit{B} at $\theta$. The resulting data is illustrated as a contour plot, with peaks and valleys in the SdH oscillations shaded in red and blue, respectively (FIG.~\ref{fig:Manuscript_FIG._2}a). As $\theta$ is rotated from $90\degree$ to 0$\degree$, their amplitude decreases and their positions shift to larger \textit{B}, indicating an increase in the cross-section area of the Fermi surface normal to \textit{B}~\cite{shoenberg2009magnetic} indicating that the contour plot of SdH oscillations serves as a signature of the closed anisotropic 3D Fermi surface in PHOPG. This characteristic is also reflected in the angle-dependent features of $R_{xy,\theta}(B)$, denoting the asymmetric part with respect to \textit{B} at $\theta$ of the Hall resistance. $R_{xy,\theta}(B)$ is the largest at $\theta=90\degree$ and monotonically decreases nearly to zero upon rotation to $\theta=0\degree$ (FIG.~\ref{fig:Manuscript_FIG._2}b). 

For CHOPG, we similarly analyze oscillations in $V_{xx,\theta}(B)$ by evaluating its second derivative with respect to \textit{B}. Here, $V_{xx,\theta}(B)$ denotes the symmetric part with respect to \textit{B} of the longitudinal voltage at $\theta$ (as we discuss below, voltage and resistance are not trivially connected as in the case of PHOPG because the current is inhomogeneous). The oscillations are illustrated as a contour plot in FIG.~\ref{fig:Manuscript_FIG._2}c, where peaks and valleys are highlighted in red and blue, respectively. A comparison between FIG.~\ref{fig:Manuscript_FIG._2}c and FIG.~\ref{fig:Manuscript_FIG._2}a reveals angle-independent features (AIFs) at specific values of \textit{B} superimposed with the SdH oscillations. For instance, the peaks at 3.5~T across all $\theta$ constitute an AIF. This AIF also manifests in $V_{xy,\theta}(B)$ (e.g., at $B=3.6$~T in FIG.~\ref{fig:Manuscript_FIG._2}d), representing the asymmetric part with respect to \textit{B} of the Hall voltage at $\theta$. Furthermore, $V_{xy,\theta}(B)$ decreases upon rotation from its maximum at $\theta=0\degree$, closely resembling $R_{xy,90\degree}$, to nearly zero for $\theta=90\degree$.

The observation of very similar SdH oscillations in $V_{xx,\theta}$ of CHOPG (alongside the additional AIF discussed below) implies that the cross-section area of the Fermi surface explored by the charge carriers in the infinitesimal strip at $\theta$ in CHOPG closely resembles that in PHOPG at $\theta$. Following the definition of $\theta$ for PHOPG (FIG.~\ref{fig:Manuscript_FIG._1}e), CHOPG will in the following be approximated as (ideally infinitely-) narrow strips of PHOPG with their \textit{c} axes oriented along the radially outward direction (see FIG.~\ref{fig:Manuscript_FIG._1}c).

\section{Modeling and interpretation of magnetotransport in curved cylinders}
\label{section4}

\begin{figure*}[ht]
\centering
\includegraphics[width=\linewidth]{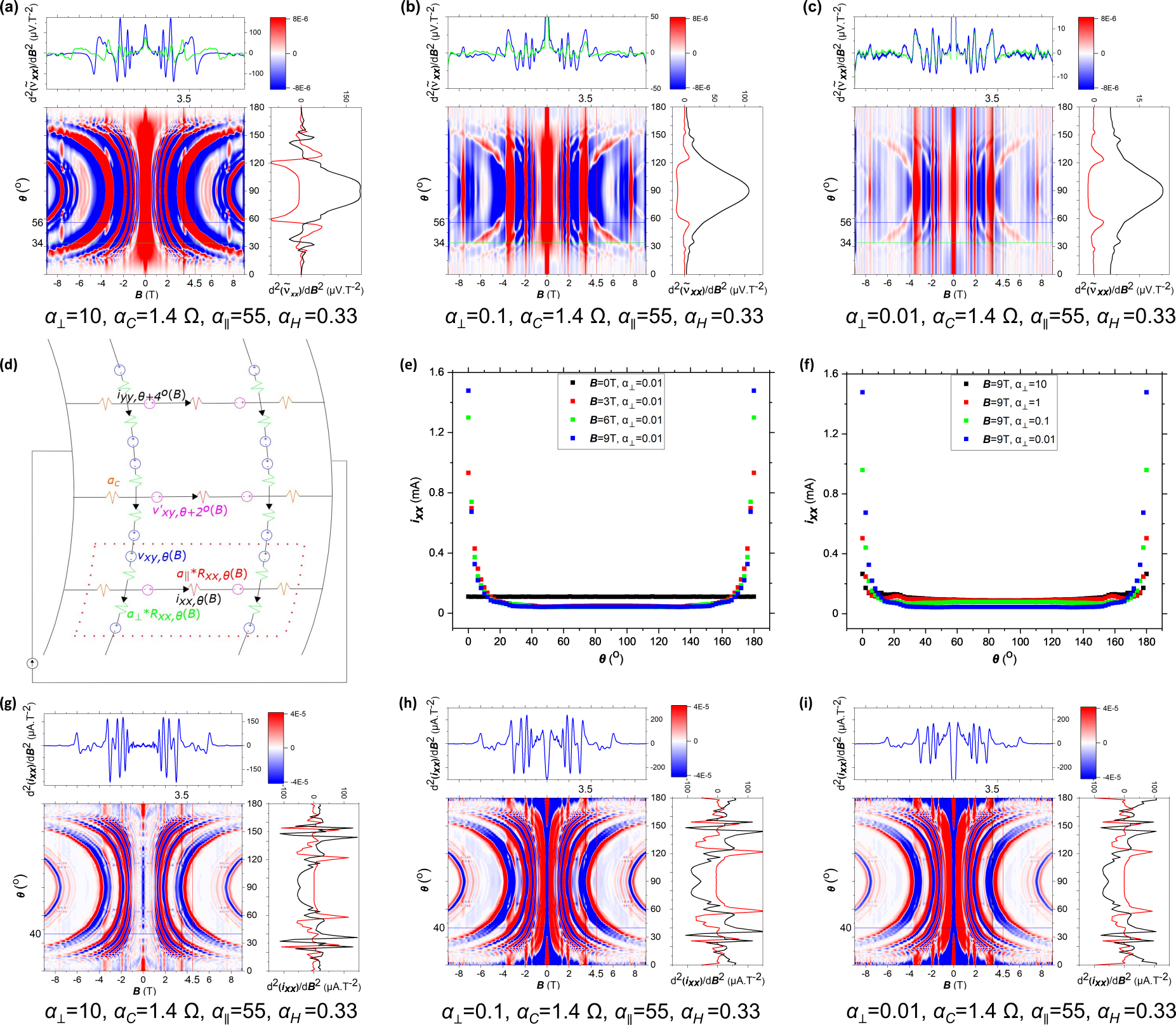}
\caption{(a-c) Contour plots of $d^2\widetilde{\mathcal{V}}_{xx,\theta}(B)/dB^2$ for $\alpha_{\perp}$ = 10, 0.1, 0.01 respectively  (where $\widetilde{\mathcal{V}}_{xx,\theta}(B)$ is the symmetric part with respect to \textit{B} of the voltage measured across the \textit{central resistor}) at $\theta$. When $\alpha_{\perp}$ is 10 the SdH oscillations are the most prominent features. For $\alpha_{\perp}$= 0.1 and 0.01, the AIF emerges in superposition with the SdH oscillations. (d) A section of the resistor network model that represents CHOPG as strips of PHOPG \textit{connected} around the circumference. (e,f) $i_{xx}$ obtained from simulations. $i_{xx}$ redistributes around the cylinder in \textit{B} and the extent of this redistribution increases with an increase in \textit{B} and decrease in $\alpha_{\perp}$. (g-i) Contour plots of $d^2i_{xx,\theta}(B)/dB^2$ for $\alpha_{\perp}$ = 10, 0.1, 0.01, respectively. Two features are observed in superposition, First, oscillations with polarity reversed with respect to the SdH oscillations. Second, additional peaks for instance, roughly between 56$\degree$ and 34$\degree$ at 3.5~T.}
\label{fig:Manuscript_FIG._3}
\end{figure*}

Based on our experimental findings, we model the charge transport on a cylindrical geometry in \textit{B} by a network model (depicted in FIG.~\ref{fig:Manuscript_FIG._3}d). Fundamentally, the model comprises PHOPG strips at $\theta \in [0\degree,180\degree$] \textit{connected} with its neighbors at $\theta \pm 2\degree$ (the network model is constructed with 2$\degree$ increments to $\theta$). This model is implemented numerically in the Simulink extension to Matlab. It takes as inputs $R_{xx,\theta}(B)$ and $R_{xy,\theta}(B)$ obtained from measurements in PHOPG, and incorporates four free parameters: $\alpha_{\parallel}$, $\alpha_{\perp}$, $\alpha_{H}$ and $\alpha_{C}$. $\alpha_{\perp}$ is the geometrical prefactor for the \textit{connecting resistor} ($\alpha_{\perp}$*$\textit{R}_{xx,\theta}(B)$) represented in green. This terminology is adopted because the resistor \textit{connects} the strip at $\theta$ to its neighbors at $\theta \pm 2\degree$. The current flowing along it is denoted as $i_{yy,\theta}(B)$. The \textit{central resistor} ($\alpha_{\parallel}$*$\textit{R}_{xx,\theta}(B)$) illustrated in red, signifies the resistance of the strip at $\theta$. $\alpha_{\parallel}$ is the geometrical prefactor that scales $\textit{R}_{xx,\theta}(B)$ measured in PHOPG to the value of the \textit{central resistor} at $\theta$. Assuming equal longitudinal resistivity in both cases, $\alpha_{\parallel}$ is determined as 110 when the strip's width is defined by the cylinder's radius multiplied by $2\degree*(\pi/180)$, and its length and thickness match the distance between longitudinal contacts and the thickness in the CHOPG sample, respectively. The current flowing along the \textit{central resistor} at $\theta$ is denoted as $i_{xx,\theta}(B)$. The voltage across the \textit{central resistor} is given by $i_{xx,\theta}(B)$*$\alpha_{\parallel}$*$\textit{R}_{xx,\theta}(B)$ and is represented by $\widetilde{\mathcal{V}}_{xx,\theta}(B)$. Orange-colored contact resistances $\alpha_{C}$ across each strip are connected to a common terminal at either end to simulate the transport measurement in CHOPG with homogeneous current contacts. 

To incorporate the Hall effect, voltage sources are introduced, as the Hall voltage is proportional to the current but develops perpendicular to the current direction. The voltage source $\nu_{xy,\theta}$=$\alpha_{H}$*$\textit{R}_{xy,\theta}(B)$*$i_{xx,\theta}(B)$ (in blue) and $\nu^{,}_{xy,\theta}$= $\alpha_{H}$*$\textit{R}_{xy,\theta}(B)$*$i_{yy,\theta}(B)$ (in pink) account for the Hall voltage due to $i_{xx,\theta}$ and $i_{yy,\theta}(B)$, respectively. $\alpha_{H}$ is the geometrical prefactor scaling $R_{xy,\theta}(B)$ in PHOPG to $\nu_{xy,\theta}$ and $\nu^{,}_{xy,\theta}$. It is determined as 0.33 based on the ratio of the thickness of CHOPG and PHOPG, assuming equal Hall resistivity in both cases.

To delve into the origin of the AIF in the contour plot of $d^2V_{xx,\theta}(B)/dB^2$, we draw attention to two observations in the contour plots of $d^2\widetilde{\mathcal{V}}_{xx,\theta}(B)/dB^2$ (FIG.~\ref{fig:Manuscript_FIG._3}a-FIG.~\ref{fig:Manuscript_FIG._3}c). (1) For $\alpha_{\perp}=10$, the contour plot closely resembles the SdH oscillations in $d^2R_{xx,\theta}(B)/dB^2$ of PHOPG (FIG.~\ref{fig:Manuscript_FIG._3}a). Faint features spanning a narrow range in $\theta$ are observed in a superposition where the amplitude of the angle-dependent features is weak, such as around $\theta=56\degree$ at 3.5~T. (2) As $\alpha_{\perp}$ is decreased to 0.1 and 0.01, the amplitude of the SdH oscillations progressively diminishes, and the AIFs emerge in superposition. The influence of $\alpha_{\perp}$ on the contour plots of $d^2\widetilde{\mathcal{V}}_{xx,\theta}(B)/dB^2$ is elucidated by discussing the contributions from each term in equation~\ref{equMain}.

\begin{widetext}
\begin{equation}
\centering
\frac{d^2\left(i_{xx,\theta}(B)*(\alpha_{\parallel}*R_{xx,\theta}(B)) \right)}{dB^2}= 
\alpha_{\parallel}*\left(i_{xx,\theta}(B)\frac{d^2R_{xx,\theta}(B)}{dB^2}+R_{xx,\theta}(B)\frac{d^2i_{xx,\theta}(B)}{dB^2} 
+2\frac{di_{xx,\theta}(B)}{dB}\frac{dR_{xx,\theta}(B)}{dB}\right)
\label{equMain}
\end{equation}
\end{widetext}

The SdH oscillations in $d^2R_{xx,\theta}/dB^2$ determine the sign of the oscillations in the first term. Its prefactor- $i_{xx}$ is uniform around the cylinder for $B=0$~T (FIG.~\ref{fig:Manuscript_FIG._3}e) since the resistance of the strips is equal for all $\theta$. However, under finite \textit{B}, $i_{xx}$ decreases from strips around $\theta=90\degree$ and increases around lower $\theta$ (FIG.~\ref{fig:Manuscript_FIG._3}f). This variation implies a redistribution of current around the cylinder, which becomes more pronounced with increasing \textit{B} and decreasing $\alpha_{\perp}$. Physically, the current redistribution arises due to the fact that the combination of local geometry and applied fields induces locally varying transport properties, even in samples that conducted homogeneously at zero field. Current flow occurs preferentially along paths with smaller resistances, determined by the smaller values of \textit{central resistors}. The \textit{central resistors}, as inferred from $R_{xx}$ of PHOPG, exhibit a monotonic dependence, decreasing as $\theta$ decreases from $90\degree$ to $0\degree$ at a fixed \textit{B} and increasing with \textit{B} rising from 0~T to 9~T at a fixed $\theta$ (refer to SI). A decrease in $\alpha_{\perp}$ further promotes the redistribution since the value of the \textit{connecting resistors} decreases. 

The sign of oscillations in the second term of equation~\ref{equMain} is governed by $d^2i_{xx,\theta}(B)/dB^2$, illustrated in FIGs.~\ref{fig:Manuscript_FIG._3}(g-i) for $\alpha_{\perp}=10$, 0.1 and 0.01, respectively. The peaks are shaded in red, and the valleys are in blue. Two features are observed in superposition: First, angle-dependent oscillations with their sign reversed compared to the SdH oscillations in $d^2R_{xx,\theta}(B)/dB^2$ (compare FIG.~\ref{fig:Manuscript_FIG._3}(g-i) with FIG.~\ref{fig:Manuscript_FIG._2}a). The sign is reversed since the \textit{central resistors} (in series with $\alpha_{C}$) are connected in parallel which enforces an inverse relation between the current ($i_{xx,\theta}(B)$) and the value of the \textit{central resistor} at $\theta$. Second, features in superposition at specific \textit{B} values arise from the redistribution of current around the cylinder. Notably, the peaks at 3.5~T roughly between $56\degree$ and $34\degree$ are significant, highlighting the dominance of current redistribution over weaker angle-dependent oscillations.

The central role played by the pronounced redistribution of current at specific \textit{B} values is crucial in observing the AIF as evident in CHOPG. For $\alpha_{\perp}=10$, where there is minimal current redistribution around the cylinder, $d^2\widetilde{\mathcal{V}}_{xx,\theta}(B)/dB^2$ closely resembles the SdH oscillations in PHOPG, akin to measurements performed with a constant current injected for all $\theta$. The faint peaks at 3.5~T around $\theta=56\degree$ can be attributed to the small but finite current redistribution. A decrease in $\alpha_{\perp}$, leads to an increase in the current redistribution, causing $i_{xx}$ (the prefactor of the first term in equation~\ref{equMain}) to drop for $\theta$ roughly between 90$\degree$ to 10$\degree$ (see FIG.~\ref{fig:Manuscript_FIG._3}f). This results in a decreased amplitude of the SdH oscillations in $d^2\widetilde{\mathcal{V}}_{xx,\theta}(B)/dB^2$ (strictly though, it only decreases for $\theta$ for which $i_{xx}$ decreases) and the signal from the redistribution of current dominates the first term across broader regions. This is observed in FIG.~\ref{fig:Manuscript_FIG._3}b and FIG.~\ref{fig:Manuscript_FIG._3}c at 3.5~T, roughly between $56\degree$ and $34\degree$, where the peaks from the redistribution of current overshadow the SdH oscillation, revealing the AIF at special \textit{B} values. 

Further reduction in $\alpha_{\perp}$ increases the extent of current redistribution, causing $\widetilde{\mathcal{V}}_{xx}(B)$ across \textit{central resistors} for all $\theta$ to converge to the same value. In the extreme case when $\alpha_{\perp}=0$, the connecting resistors behave like ideal wires, and only the AIF is observed (refer to SI).

The third term in equation~\ref{equMain}, finally, smears the position in $B$ of peaks and valleys in $d^2\widetilde{\mathcal{V}}_{xx,\theta}(B)/dB^2$ with respect to the SdH oscillation of PHOPG since it contains first derivatives in contrast to the
second derivatives in the first and second terms of equation~\ref{equMain} (refer to plots in the SI).

To elucidate the origin of the AIF in the Hall measurements, the sign of  $\nu_{xy,\theta}$ is initially assigned to the model by evaluating the cross product of $i_{xx}$ and the component of \textit{B} directed normal to the surface of each strip. Thus for \textit{B} directed along \textit{z} and $i_{xx}$ directed into the plane, $\nu_{xy,\theta}$ is positive in the
anti-clockwise direction for all strips with $\theta$ $\in$ (0$\degree$,180$\degree$) and negative for $\theta$ $\in$ (180$\degree$,360$\degree$). Equation~\ref{equ1} is then constructed to evaluate the Hall voltage at $\theta$, denoted by $\widetilde{\mathcal{V}}_{xy,\theta}$ since this value cannot be directly obtained from the model by measuring the Hall voltage across the strips at $\theta$ and 180$\degree$+$\theta$ as the model is constructed only for $\theta$ $\in$ [0$\degree$,180$\degree$]. Equation~\ref{equ1} reveals that, $\widetilde{\mathcal{V}}_{xy,\theta} (B) = \sum_{\phi=\theta}^{\phi=180\degree+\theta} \nu_{xy,\phi}(B)$.

\begin{equation}
\centering
\widetilde{\mathcal{V}}_{xy,\theta}= \widetilde{\mathcal{V}}_{xy,0\degree}-\widetilde{\mathcal{V}}_{xy,\theta-0\degree}-\widetilde{\mathcal{V}}_{xy,180\degree-(180\degree-\theta)}  
 \label{equ1}
\end{equation}

The first, second, and third terms denote the asymmetric part with respect to \textit{B} of the Hall voltage measured in the model between the strips at 0$\degree$, 180$\degree$ and $\theta$, 0$\degree$ and 180$\degree$, 180$\degree$-$\theta$, respectively. The first term is analogous to $V_{xy,0\degree}$ in CHOPG. The second term is subtracted to account for the difference between the dashed semicircle and the semicircle between 0$\degree$ and 180$\degree$. The third term accounts for the additional segment in the dashed semicircle between 180$\degree+\theta$ and 180$\degree$. By symmetry (about the \textit{xy} plane) its magnitude is equal to the asymmetric part with respect to \textit{B} of the voltage measured between the stripes at 180$\degree$ and 180$\degree$-$\theta$. However, since its sign is opposite, this term is subtracted. 

\begin{figure}[H]
\centering
\includegraphics[width=\linewidth]{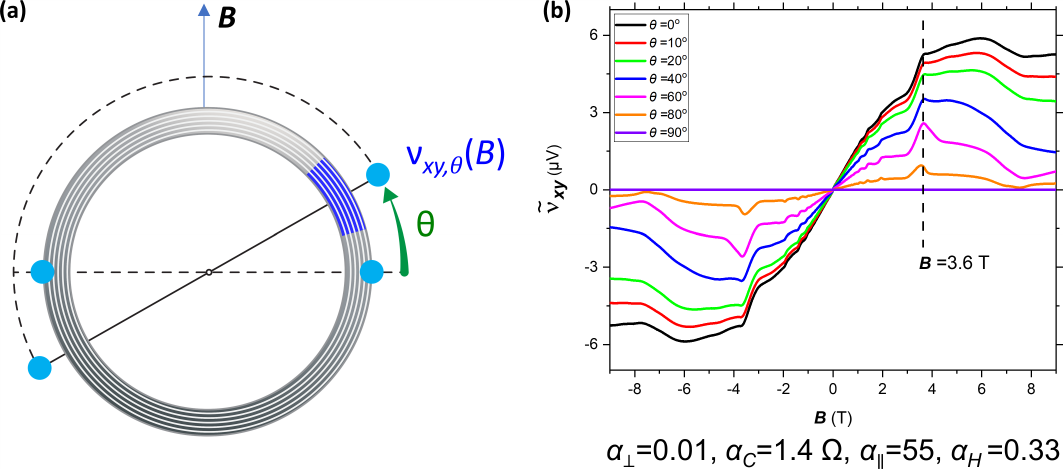}
\caption{(a) Front view of the cylinder depicting $\nu_{xy,\theta}$ across the strip at $\theta$ due to $i_{xx,\theta}(B)$ directed into the plane and \textit{B} directed along \textit{z} direction. (b) $\widetilde{\mathcal{V}}_{xy,\theta}$ obtained from simulations for $\alpha_{\perp}=0.01$ reproduces the AIF.}
\label{fig:Manuscript_FIG._4}
\end{figure}

The results of the $\widetilde{\mathcal{V}}_{xy,\theta}$ plotted in FIG.~\ref{fig:Manuscript_FIG._4}b evaluated for $\alpha_{\perp}=0.01$ show reasonable quantitative agreement with $V_{xy,\theta}$. The magnitude of $\widetilde{\mathcal{V}}_{xy,\theta}$ is maximum at $\theta=0\degree$ and monotonically decreases to zero for $\theta=90\degree$ since the first term in equation~\ref{equ1} is exactly canceled by the sum of the second and third terms. The AIFs appear in $\widetilde{\mathcal{V}}_{xy,\theta}$ since it includes a sum of $\nu_{xy,\theta}$ over several strips around $\theta=90\degree$, which exhibit pronounced Hall features around 3.6~T for instance, see FIG.~\ref{fig:Manuscript_FIG._2}c and FIG.~\ref{fig:Manuscript_FIG._3}i.
\\

\section{Conclusions}
\label{section5}

To summarize, we performed measurements of magnetotransport in 3D graphite samples. We analyzed the longitudinal and Hall responses of both planar slabs and cylinders. Based thereon, we developed a simple phenomenological resistor network model. This model was shown to describe the observed magnetotransport in 3D graphite cylinders both qualitatively and quantitatively. We find that the combination of a non-trivial, curved sample geometry and an applied homogeneous magnetic field induces locally varying transport properties depending on the relative orientation of the local sample normal and magnetic field. The locally varying transport properties in turn lead to a non-trivial redistribution of current across the sample as the magnetic field changes. As a result, magnetotransport in cylindrical samples shows a superposition of features that depend on the relative angle between the magnetic field and surface normal along the current path (``angle-dependent features''), and angle-independent features. While the former can be understood by direct analogy to planar graphite slabs at varying angles, the latter represents the tendency towards an angle-independent voltage drop determined by the parallel resistance of all sample regions. This tendency arises from effective metallic links perpendicular to the current path between different cylinder segments. Angle-independent features in the magnetoresistance of curved 3D systems are thus a direct consequence of the absence of a bulk gap in 3D systems subject to magnetic fields -- a stark contrast to curved 2D samples in strong fields.

While we focused on cylindrical samples as a well-defined reference case for curved systems, we expect our findings to qualitatively carry over to other sample geometries and materials. In particular, the phenomenological network model developed in this work can straightforwardly be adapted to other scenarios. Our findings therefore pave the way for transport studies in three-dimensional curved samples on a broad basis.  

\begin{acknowledgments}
TM acknowledges funding by the Deutsche Forschungsgemeinschaft (DFG) via the Emmy Noether Programme (Quantum Design grant, ME4844/1, project-id 327807255), project A04 of the Collaborative Research Center SFB 1143 (project-id 247310070), and the Cluster of Excellence on Complexity and Topology in Quantum Matter ct.qmat (EXC 2147, project-id 390858490), as well as from the Luxembourg National Research Fund (FNR) and the DFG through the CORE grant “Topology in relativistic semimetals (TOPREL)” (FNR project No. C20/MS/14764976 and DFG project-id 452557895).
\end{acknowledgments}

\bibliography{main.bib}
\bibliographystyle{unsrt}

\newpage

\section{Supporting information}

\subsection{Comparing the longitudinal resistance in PHOPG with the longitudinal voltage in CHOPG}

\begin{figure*}[ht]
\centering
\includegraphics[width=\linewidth]{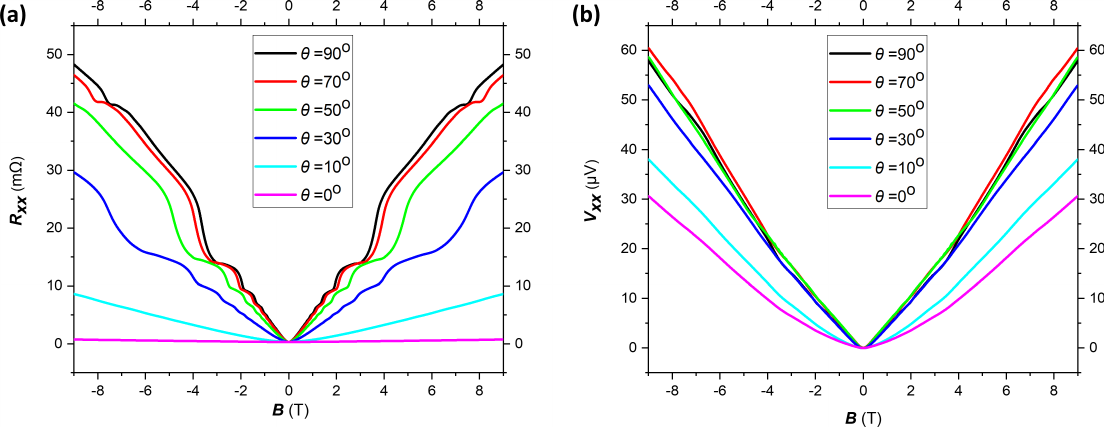}
\caption{(a) $R_{xx}(B)$ in PHOPG (b) $V_{xx}(B)$ in CHOPG at various $\theta$ values.}
\label{fig:Manuscript_SI1}
\end{figure*}

$R_{xx,\theta}(B)$, the symmetric part \textit{w.r.t.} \textit{B} of the magnetoresistance at $\theta$ in PHOPG, evaluated from the ratio of longitudinal voltage and current (\textit{i}), is the largest at $\theta=90\degree$ and decreases monotonically upon rotation to $\theta=0\degree$ (FIG.~\ref{fig:Manuscript_SI1}a). This monotonic dependence is absent in $V_{xx,\theta}(B)$, which denotes the symmetric part \textit{w.r.t.} \textit{B} of the longitudinal voltage at $\theta$ in CHOPG (FIG.~\ref{fig:Manuscript_SI1}b). Possible reasons for this non-monotonic dependence are elaborated below. 

\subsection{A comparison among various CHOPG samples}

\begin{figure*}[ht]
\centering
\includegraphics[width=\linewidth]{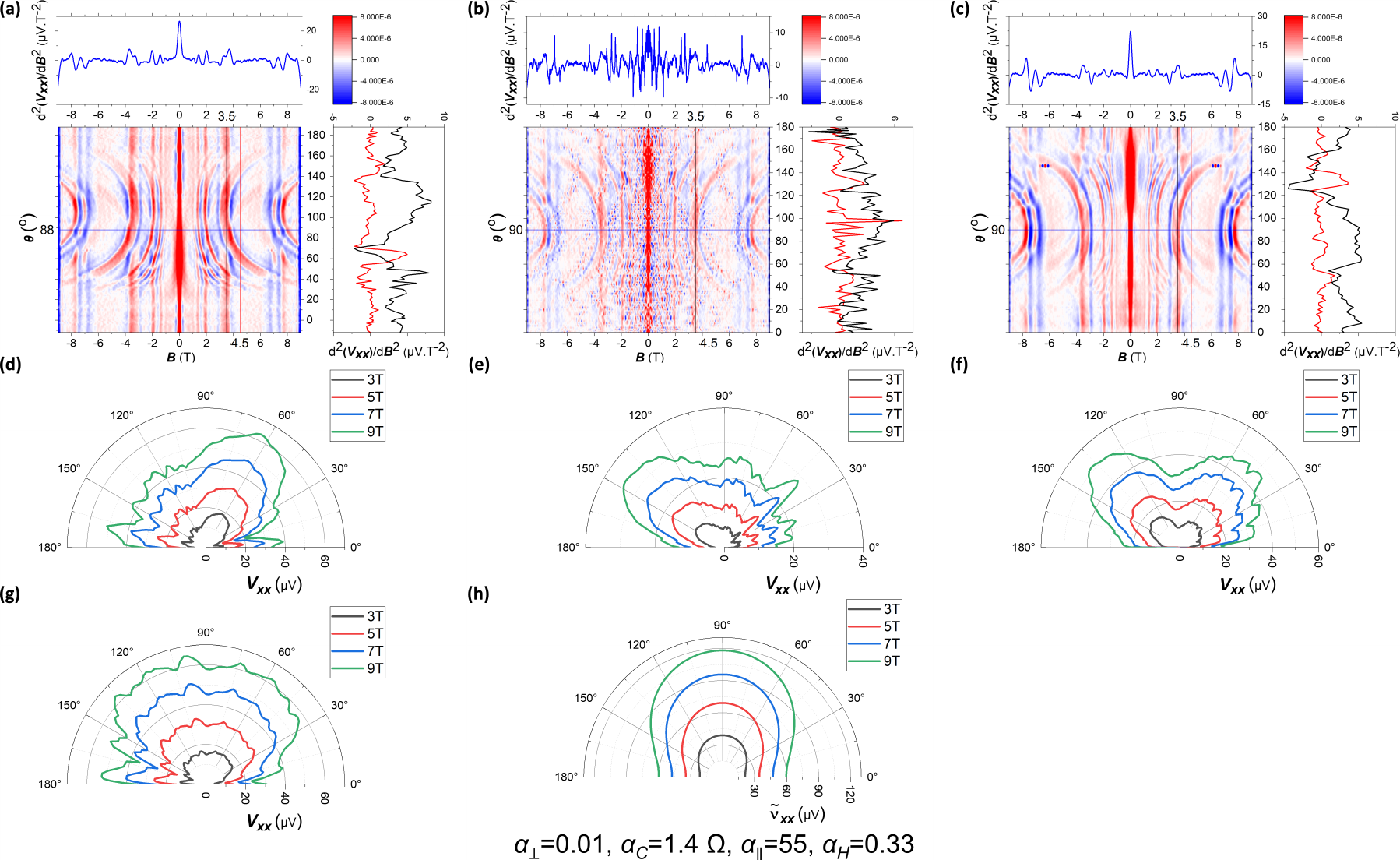}
\caption{Contour plots of $d^2V_{xx,\theta}(B)/dB^2$ for (a), (b) both pairs of contacts in CHOPG S2 and (c) across a pair of contacts in CHOPG S3. (d-f) Corresponding polar plots of $V_{xx}$ at various values of $B$ against $\theta$. (g) Polar plot of $V_{xx}$ for CHOPG S1. (h) Polar plot of $\widetilde{\mathcal{V}}_{xx}(B)$ obtained from simulating the network model.}
\label{fig:SI_Repeatsonrotationarounsaxis}
\end{figure*}

FIGs~\ref{fig:SI_Repeatsonrotationarounsaxis}(a-c) illustrate the contour plots of $d^2V_{xx,\theta}(B)/dB^2$ derived from measurements conducted on two additional CHOPG samples (S2 and S3). These plots display the AIF in superposition with the SdH oscillations. The corresponding $V_{xx}$ values of FIGs~\ref{fig:SI_Repeatsonrotationarounsaxis}(a-c) are represented as polar plots in FIGs.~S\ref{fig:SI_Repeatsonrotationarounsaxis}(d-f), portraying the dependence on $\theta$ at various values of $B$. FIG.~\ref{fig:SI_Repeatsonrotationarounsaxis}g presents the polar plot for CHOPG S1, whose contour plot is discussed in the main manuscript. 

The polar plots reveal a sample-dependent non-monotonic behavior, in contrast to the monotonic trend of $\widetilde{\mathcal{V}}_{xx}$ observed in FIGs~\ref{fig:SI_Repeatsonrotationarounsaxis}(h). A non-monotonic dependence in $\widetilde{\mathcal{V}}_{xx}$ can be introduced in the model by selecting different values of $\alpha_C$ for strips across $\theta$, suggesting a non-uniform current injection around the circular rim of CHOPG. Furthermore, varying values of $\alpha_\parallel$ for strips across $\theta$ imply an inequality in the longitudinal resistivity of the strip in CHOPG compared to PHOPG, even for identical $\theta$ values. Physically, this may stem from intrinsic differences among the strips in CHOPG, potentially arising due to variations in defect density and thickness around the cylinder introduced during the fabrication process. Random defects and impurities may influence the value of the \textit{central resistor} without severely distorting the SdH oscillations.

\subsection{The network model for $\alpha_\perp=0$}

\begin{figure*}[ht]
\centering
\includegraphics[scale=1.2]{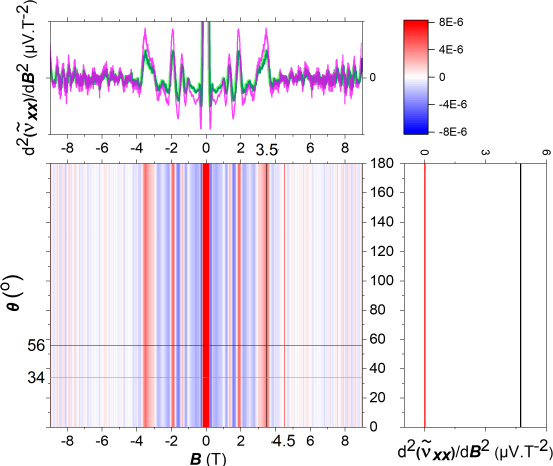}
\caption{Contour plot of $d^2\widetilde{\mathcal{V}}_{xx,\theta}(B)/dB^2$ simulated for $\alpha_{\perp}=0$. Overlaid in pink are oscillations in the value of the effective resistor, showcasing good alignment with the AIF.}
\label{fig:alpha_perp_zero}
\end{figure*}

In the limit where $\alpha_{\perp}=0$, the \textit{connecting resistors} transform into ideal wires, causing $\widetilde{\mathcal{V}}_{xx}(B)$ to precisely coincide at all $\theta$, forming the AIF as depicted in  FIG.~\ref{fig:alpha_perp_zero}. In this scenario, the AIF can be conceptualized as the oscillations in the value of an effective resistor. The inverse of this value is determined from the inverse sum of the values of the \textit{central resistors}, emphasizing their parallel connection. 

The oscillatory component in the effective resistor's value is evaluated from its double derivative \textit{w.r.t.} \textit{B}, illustrated in pink in FIG.~\ref{fig:alpha_perp_zero}. The agreement between the oscillations of the effective resistor's value and the AIF underscores this interpretation of the AIF. Essentially, the AIF can be perceived as reflective of the oscillations in the value of this effective resistor.

\subsection{Contribution from the third term}

\begin{figure*}[ht]
\centering
\includegraphics[width=\linewidth]{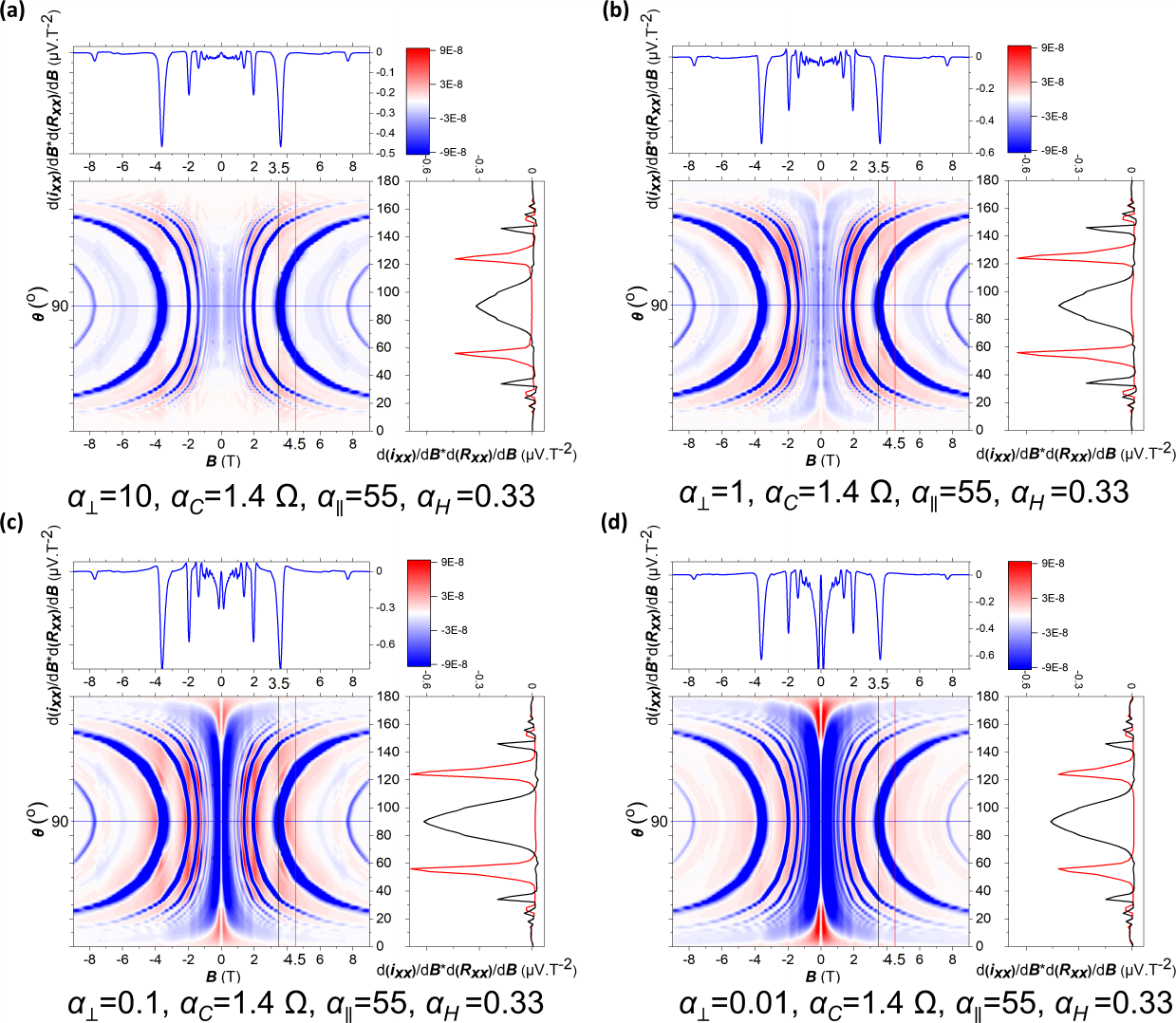}
\caption{Contour plots of $(di_{xx,\theta}(B)/dB)*(dR_{xx,\theta}(B)/dB)$ for (a-d) $\alpha_{\perp}$=10, 1, 0.1 and 0.01, respectively, and $\alpha_{C}$=1.4, $\alpha_{\parallel}$=55, $\alpha_{H}$=0.33}
\label{fig:SI_dIdR}
\end{figure*}

FIGs.~S\ref{fig:SI_dIdR}(a-d) present contour plots of $(di_{xx,\theta}(B)/dB)*(dR_{xx,\theta}(B)/dB)$ at various $\alpha_\perp$ values, illustrating the contribution of this term to the expansion of $\frac{d^2(i_{xx,\theta}(B)*(\alpha_{\parallel}*R_{xx,\theta}(B)))}{dB^2}$. As $(di_{xx,\theta}(B)/dB)*(dR_{xx,\theta}(B)/dB)$ involves only first derivatives, the positioning of peaks and valleys in its contour plots undergoes a shift when compared to the oscillations in $d^2R_{xx,\theta}(B)/dB^2$ and $d^2i_{xx,\theta}(B)/dB^2$. Consequently, this shift results in the smearing of the positions in \textit{B} where oscillations are observed in $d^2\widetilde{\mathcal{V}}_{xx,\theta}(B)/dB^2$ in relation to the SdH oscillations.

\end{document}